\documentstyle[preprint,aps]{revtex}

\begin{document}

\draft

%\preprint{xxx-xx/95}

\title{Exotic Solutions in Einstein-Antisymmetric Tensor Theory}

\author{Hongsu Kim}

\address{Department of Physics\\
Sogang University, C.P.O. Box 1142, Seoul 100-611, KOREA}

\date{August, 1995}

\maketitle

\begin{abstract}
Classical solutions of the self-interacting, non-abelian antisymmetric tensor
gauge theory of Freedman and Townsend coupled to Einstein gravity is discussed.
Particularly, it is demonstrated that the theory admits a classical metric
solution which, depending on the value of the gauge coupling parameter of
the theory, exhibits a black hole with an exotic non-abelian hair 
or a spacetime showing the ``violation of the cosmic
censorship hypothesis" which should be distinguished from white holes. 
\end{abstract}


\narrowtext
%\twocolumn

      In this letter, we consider the classical solutions of the antisymmetric 
tensor gauge theory coupled to the Einstein gravity. Generally, the
antisymmetric tensor field occurs in some formulations of supergravity theories,
relativistic strings, etc[1-3]. In particular, a model of a self-interacting
antisymmetric tensor field possessing a non-abelian gauge invariance was first
discussed by Freedman and Townsend (FT)[4] in which they established, among
other things, the on-shell equivalence between the antisymmetric tensor gauge
theory and the non-linear sigma model. Subsequent works attempting the
quantizaion of the self-interacting antisymmetric tensor theory of FT also have
appeared in the literature [5].
As for the motivation for the present work, we are interested in a rather 
different aspect of the theory; we look for the classical metric and matter
solutions of the ``Einstein-Antisymmetric tensor (EAT)" theory. As the main
result, we will demonstrate that the theory admits a classical metric solution
which, depending on the value of the gauge coupling parameter of the theory, 
exhibits a black hole with a non-abelian hair
or a spacetime showing the ``violation of the cosmic censorship hypothesis."
Before formualting our theory, we begin by briefly reviewing the antisymmetric 
tensor gauge theory in flat Minkowski spacetime [4].
The self-interacting, non-abelian
antisymmetric tensor gauge theory of FT can be formulated in its second-order
version or alternatively in its first-order version [4,5].
And the two formulations are, of course, on-shell equivalent.
For some purposes, the ``first-order formulation" of the theory is 
convenient. In this formulation, the action is given by
\begin{eqnarray}
L={ 1 \over 8} \epsilon ^{\mu \nu \rho \sigma} B^a_{\mu \nu}F^a_{\rho \sigma}
  - {1 \over 8}A^a_{\mu}A^{a \mu}
\end{eqnarray}
where the antisymmetric tensor gauge field $B^a_{\mu\nu}$ and a vector potential 
$A^a_{\mu}$ associated with the field strength $F^a_{\mu\nu}$ are treated as being
independent variables of the theory.
Now by varying this Lagrangian with respect to these independent variables, we can 
obtain the classical field equation for $B^a_{\mu \nu}$ and $A^a_{\mu}$
respectively as 
\begin{eqnarray}
& &F^a_{\mu\nu}=0, \\
& &D^{\mu}\tilde{B}^a_{\mu\nu}+A^a_{\nu}=0
\end{eqnarray} 
\begin{eqnarray}
\rm{where}~~ & &F^a_{\mu\nu}=\partial_\mu A^a_\nu - \partial_\nu A^a_\mu 
+ g_c f^{abc}A^b_\mu A^c_\nu, \nonumber \\
& &D^{ac}_\mu = (\partial_\mu \delta ^{ac}+g_c f^{abc}A^b_{\mu}),~~
\rm{and}~~\tilde{B}^{a}_{\mu\nu}=\epsilon_{\mu\nu}^{~~\rho\sigma}B^{a}_{\rho\sigma}.
\nonumber 
\end{eqnarray}  
Further, by acting the operator $D^\nu$ on the field
equation for $A^a_\mu$ in eq.(3) and using eq.(2), we get 
\begin{eqnarray}
D^\mu A^a_\mu =0.
\end{eqnarray}
Then the vector potential $A^a_\mu$ satisfying classical field equation eqs.(2) 
and (3) turns out to be the `pure gauge',
\begin{eqnarray}
A_\mu=-{i \over g_c}(\partial_\mu U(x))U^{-1}(x)
\end{eqnarray}
(where $U(x)$ is the gauge tranformation function of the given non-abelian 
gauge group) provided it (eq.(5)) satisfies the ``combined" field equation (4).
Using the classical field equation for $A^a_\mu$ in eq.(3) one can show that 
the first-order formulation Lagrangian in eq.(1) turns into the second-order
formulation Lagrangian [4]. Further, since the classical field equation for
$B^a_{\mu\nu}$ is the vanishing $F^a_{\mu\nu}$, in this first-order formulation
$B^a_{\mu\nu}$ field appears classically as a Lagrange multiplier enforcing
the constraint $F^a_{\mu\nu} = 0$. And finally, substituting the pure
gauge solution in eq.(5) into the Lagrangian (1) demonstrates the equivalence of
the theory to the non-linear $\sigma$-model [4].
Now we consider the case when the gravity is turned on. 
To begin, it seems essential for us to declare our sign convention.
We choose to take the convention in which $g_{\mu\nu} = 
{\rm diag}(- + + +)$ and
$R^{\alpha}_{\beta \mu \nu} = \partial_{\mu}\Gamma^{\alpha}_{\beta \nu} -
\partial_{\nu}\Gamma^{\alpha}_{\beta \mu} + \Gamma^{\alpha}_{\mu \lambda}
\Gamma^{\lambda}_{\beta \nu} - \Gamma^{\alpha}_{\nu \lambda}
\Gamma^{\lambda}_{\beta \mu}.$   It is crucial to fix the right sign for
the antisymmetric tensor (i.e., the matter) sector of the action ``relative"
to the gravity action. Therefore in our sign convention, we explain the
way we determined the sign for the matter action using the fact that
on-shell, the antisymmetric tensor theory action is equivalent to that of
non-linear sigma model with the right sign.
Consider generators $T^{a}$ of the non-abelian group $G$ in a representation
in which $[T^{a}, T^{b}] = i f^{abc}T^{c}$,  ${\rm Tr}(T^{a}T^{b}) = c
\delta^{ab}$ and $U(x)=\exp{[i\phi^{a}(x)T^{a}]}$ where $f^{abc}$ and $c$
are the structure constant and a representation-dependent positive constant
respectively. Then it can be readily shown that upon substituting the
on-shell condition, $F_{\mu\nu} = 0$, i.e., $A_{\mu}=-{i\over g_{c}}
(\partial_{\mu} U)U^{-1}$, one gets 
$L = ({1\over 8c}){\rm Tr}[\epsilon^{\mu\nu\rho\sigma}B_{\mu\nu}F_{\rho\sigma}
-A_{\mu}A^{\mu}] = (-{1\over 8cg^{2}_{c}}){\rm Tr}[(\partial_{\mu} U^{-1})
(\partial^{\mu} U)]$ (with $B_{\mu\nu} = B^{a}_{\mu\nu}T^{a}$ and 
$A_{\mu} = A^{a}_{\mu}T^{a}$) which is of the right sign. Thus in this way
we have fixed the sign for the matter action.
Now in order to describe 
the coupled Einstein-Antisymmetric tensor theory we again employ the 
first order
formulation of the antisymmetric tensor sector, then the theory is described by
the action (we work in the unit $G = 1$)
\begin{eqnarray} 
S &=& S_G + S_{AT} \\
  &=& \int d^4 x \sqrt{g} [ {1 \over {16 \pi }}R 
  + {1 \over 8}
\left( {1 \over \sqrt{g}}g^{\mu\alpha} g^{\nu\beta} \tilde{B}^{a}_{\mu\nu}
F^a_{\alpha\beta} - g^{\mu\nu} A^a_{\mu} A^a_{\nu} \right) ] \nonumber 
\end{eqnarray}
where we used that in curved spacetime, 
$\epsilon^{\mu\nu\rho\sigma} \rightarrow \left(\epsilon^{\mu\nu\rho\sigma} \over
\sqrt{g} \right)$  and again $\tilde B ^a_{\mu\nu} =
\epsilon_{\mu\nu}^{~~~\rho\sigma}B^a_{\rho\sigma}=g_{\mu\alpha}
g_{\nu\beta}\epsilon^{\alpha\beta\rho\sigma}B^a_{\rho\sigma}$.
The curved spacetime version of the classical field equation for 
$B^a_{\mu\nu}$ and $A^a_\mu$ are given respectively by
\begin{eqnarray}
& &F^a_{\mu\nu}=0, \\
& &D^\mu \tilde B^a_{\mu\nu}+\sqrt{g}A^a_\nu=0
\end{eqnarray}
along with the curved spacetime counterpart of the eq.(4) which is 
the necessary condition that $A^a_\mu$ must satisfy as a classical solution
being given by
\begin{eqnarray}
D^\mu \left( \sqrt{g} A^a_\mu \right)=0.
\end{eqnarray}
In addition, varying the action in eq.(6) with respect to the metric 
$g_{\mu\nu}$ yields the Einstein field equation
\begin{eqnarray}
R_{\mu\nu} = - 4 \pi [{1\over\sqrt g} \{g^{\alpha\beta}(\tilde 
B^a_{\mu\alpha} F^a_{\nu\beta}) - {1\over 2}g_{\mu\nu}(\tilde
B^a_{\alpha\beta}F^{a\alpha\beta})\}-{1\over 2}(A^a_\mu A^a_\nu)] 
\end{eqnarray}
with the energy-momemtum tensor being given by 
\begin{eqnarray}
T_{\mu\nu}= - {1\over 8}
[{1\over \sqrt{g}}g^{\alpha\beta}(4\tilde B^a_{\mu\alpha}
F^a_{\nu\beta}) 
+\{g_{\mu\nu}(A^a_\alpha A^{a\alpha})-2(A^a_\mu A^a_\nu)\} ]. \nonumber
\end{eqnarray} 
Our strategy for solving the classical equation of motion in eqs.(7),(8) and 
(10) along with the necessary condition eq.(9) is as follows ; we start with the
solution to the field equations (7) and (9) which, as we shall see, still turns
out to be the pure gauge in eq.(5) even in the curved spacetime. Next, for
this pure gauge solution satisfying
eqs.(7) and (9), the Einstein field equation in (10) takes a remarkably simple
form, $R_{\mu\nu}= 2\pi (A^a_\mu A^a_\nu )$ with
$T_{\mu\nu}={1\over4}\left[A^a_\mu A^a_\nu-{1\over2}g_{\mu\nu}\left(A^a_\alpha
A^{a\alpha}\right)\right]$ that can be readily solvable. Finally by substituting
the pure gauge solution for $A^a_\nu$ and the metric solution $g_{\mu\nu}$
into the field equation (8) (plus possibly
the gauge condition of the form $D_\mu B^{\mu\nu}=0$), one can, in principle,
obtain the classical solution for $B^a_{\mu\nu}$. We will not, however,
expilicitly solve for $B^a_{\mu\nu}$ here, partly because we are essentially
interested in the spacetime metric solution which, as mentioned, is independent
of the solution form of $B^a_{\mu\nu}$ and partly because $B^a_{\mu\nu}$ appears
classically as a Lagrange multiplier enforcing the constraint $F^a_{\mu\nu}=0$.
	Now suppose we look for static, spherically-symmetric solutions to the 
classical field equations that are also asymptotically flat. Then, first the
metric can be written in the form 
\begin{eqnarray}
ds^2=-B\left(r\right)dt^2+A\left(r\right)dr^2+r^2d\Omega^2_2
\end{eqnarray}
with $d\Omega^2_2$ being the metric on the unit two-sphere.
Next, for the matter sector, especially for the vector potential solution of 
the pure gauge form in eq. (5), in order to
look for a spherically-symmetric solution we take the standard ansatz which is
the same in form as the flat spacetime Wu-Yang monopole solution ansatz [6]
(with the non-abelian gauge group for the antisymmetric tensor field being
chosen to be SU(2))
\begin{eqnarray}
A^a_0\left(r \right) &=& 0, \nonumber\\
A^a_i\left(r \right) &=& -\epsilon_{iab} {x^b \over g_c r^2} 
\left[ 1- u\left(r \right) \right].
\end{eqnarray}
As is well-known, this solution ansatz is indeed spherically-symmetric 
in the sense that the effect of a spatial rotation, SO(3) can be compensated by
a gauge transformation, SU(2). In the spherical-polar coordinates, this ansatz
for the vector potential $A^a_\mu$ and the non-vanishing components of the
corresponding field strength $F^a_{\mu\nu}$ are given by
\begin{eqnarray}
A^a_0=A^a_r=0,
~~A^a_\theta=-{1\over g_c}[1-u(r)]\hat\phi^a, 
~~A^a_\phi={1\over g_c} [1-u(r)]\sin \theta \hat\theta^a 
\end{eqnarray}
and
\begin{eqnarray}
F^a_{r\theta}={u'(r) \over g_c}\hat\phi^a, 
~~F^a_{r\phi}= -{u'(r) \over g_c}\sin \theta \hat\theta^a,
~F^a_{\theta\phi}={[u^2(r)-1]\over g_c^2}\sin \theta \hat r^a \nonumber
\end{eqnarray}
where prime denotes the derivative with respect to $r$ and
\begin{eqnarray}
\hat r^a&=&(\sin \theta \cos \phi,\sin \theta \sin \phi,\cos \theta), \nonumber
\\
\hat \theta^a&=&(\cos \theta \cos \phi,\cos \theta \sin \phi,-\sin \theta),\nonumber \\
\hat \phi^a&=&(-\sin \phi,\cos \phi,0). \nonumber
\end{eqnarray}
Here, it is interesting to note that, the case $u(r)=0$ corresponds to 
the exact albeit singular monopole solution of Wu-Yang type with nonvanishing
$A^a_\mu$ and $F^a_{\mu\nu}$ ; the case $u(r)=+1$ corresponds to the `trivial'
vacuum solution with vanishing $A^a_\mu$ and $F^a_{\mu\nu}$ ; and finally the case
$u(r)=-1$ corresponds to a `non-trivial' vacuum solution with vanishing
$F^a_{\mu\nu}$ but non-vanishing $A^a_\mu$. 
 Therefore, since we are looking for a
non-trivial pure gauge solution satisfying $F^a_{\mu\nu}=0$, we should take the last case
with $u(r)=-1$. Further one can easily check that this non-trivial vacuum gauge
solution $A^a_0=0,A^a_i=-\epsilon_{iab}(2x^b/g_c r^2)$ does satisfy the necessary
condition that it must satisfy, $D^\mu(\sqrt{g}A^a_\mu)=0$ in eq.(9).
Now that we have established the spherically-symmetric vector potential 
solution to field equations in curved spacetime. As mentioned earlier, then, our
next job is to substitute this non-trivial vector potential solution into the
Einstein field equation in (10) to solve for the spacetime metric solution.
The resulting Einstein equation now reads,
\begin{eqnarray}
R_{\mu\nu}&=&2\pi (A^a_\mu A^a_\nu), \nonumber \\
T_{\mu\nu}&=&{1 \over 4}[A^a_\mu A^a_\nu - {1 \over 2} 
g_{\mu\nu}(A^a_\alpha A^{a\alpha})], \nonumber \\
\rm{with}~~~ A^a_{0}&=&A^a_r=0~,~~A^a_\theta=
-{2\over g_c} \hat\phi^a~,~~A^a_\phi={2 \over g_c}\sin \theta \hat \theta^a.
\nonumber
\end{eqnarray}
Note that in terms of the isotropic metric given in eq.(11) only two 
components of the Einstein equations out of the three are truely independent
because the third component is satisfied automatically due to the
energy-momemtum conservation, $T^{\mu\nu}_{~~;\mu}=0$. Thus we consider the
following two independent combinations convenient for solving the Einstein
equations,
\begin{eqnarray}
{1\over AB}(AR_{tt}+BR_{rr})=8\pi [-T^t_t +T^r_r], \nonumber \\
{1\over 2}({1\over B}R_{tt}+{1\over A}R_{rr})+{1\over r^2}R_{\theta\theta}=8\pi 
[-T^t_t].
\end{eqnarray}
The first combination yields $B(r)=A^{-1}(r)$ where we imposed the 
asymptotic flatness condition, $A(r) \rightarrow 1,B(r)\rightarrow1$ as 
$r \rightarrow \infty.$ On the other hand the second combination gives 
\begin{eqnarray}
A(r)=[1-{2M(r) \over r}]^{-1} \nonumber
\end{eqnarray}
where $M(r)$ is to be determined from 
${dM(r)\over dr}=4\pi r^2 \rho_m(r)=(4\pi/g^2_c)$ with $\rho_m(r)=[-T^t_t]$.
Namely, $M(r) = M + 4\pi r/g^2_{c}$ with the integration constant $M$ being
identified with the total mass-energy of the system defined at the spatial
infinity, $i^{0}$, namely the ``ADM mass".
 Finally, the classical vector potential and the metric solution are given by
\begin{eqnarray}
A &=& A_\mu dx^\mu={1\over g_c}
[-2\tau_\phi d\theta + 2\sin\theta\tau_\theta d\phi], \\
d s^2 &=& -[(1-{8\pi \over g_c^2})-{2M \over r}]dt^2 +
[(1-{8\pi \over g_c^2})-{2M \over r}]^{-1}dr^2+r^2 d\Omega^2_2 \nonumber 
\end{eqnarray}
where 
$\tau_r \equiv \hat r^a (\sigma^a/2),\tau_\theta = \hat \theta^a( \sigma^a/2),
\tau _\phi=\hat \phi^a(\sigma^2/2)$ with $\sigma^a$ being the Pauli spin
matrices. 
Note that there is also a trivial vacuum solution with corresponding 
gauge
potential and metric being given by $A_{\mu} = 0$ (or $u(r) = +1$) and
the usual Schwarzschild solution respectively. Here it is interesting to
recognize that although the two gauge potential solutions, trivial vacuum
$A_{\mu} = 0$ and the nontrivial vacuum gauge $A_{\mu} = (-i/g_{c})
(\partial_{\mu}U)U^{-1}$, are related by a gauge transformation and hence
produce the same field strength tensor $F_{\mu\nu} = 0$, the spacetime
metrics generated by each of the two gauge choices above are not related
by any coordinate transformation and thus produce distinct curvatures.
This can be easily seen by evaluating the curvature invariant
$I = R_{abcd}R^{abcd}$ with $a$,$b$,$c$,$d$ being indices associated with
an orthonormal basis. In the same Schwarzschild coordinates, the curvature
invariant of the usual Schwarzschild solution is given by
$I = {48M^{2}\over r^{6}}$ whereas that of the metric solution in eq.(15) 
turns
out to be $I = 16[2+(1+{4\pi r\over g^{2}_{c}M})^{2}]{M^{2}\over r^{6}}$.
Now, we would like to examine the nature of the spacetime described by 
our metric solution in eq.(15). To do so we consider three cases :
\\
(i) In the weak coupling limit ($g_{c} << 2\sqrt{2\pi}$) ;
\begin{eqnarray}
ds^2 = [({8\pi\over g^{2}_{c}} - 1) + {2M\over r}]dt^{2}
- [({8\pi\over g^{2}_{c}} - 1) + {2M\over r}]^{-1}dr^{2}
+ r^{2}d\Omega^{2}_{2}. \nonumber
\end{eqnarray}
This metric represents a spacetime in which $r$ is timelike and $t$ is
spacelike. Thus the metric has an explicit time-dependence. The curvature
singularity at $r = 0$ is timelike and the future of any Cauchy surface
contains a naked singularity which is visible from the future null
infinity $I^{+}$. Namely no event horizon arises and thus it exhibits
an example of the violation of cosmic censorship hypothesis.
Another peculiar characteristic of this spacetime is that when one
examine its timelike geodesics, one finds that they involve {\it attractive}
centrifugal potentials rather than repulsive barriers.
\\
(ii) For the coupling constant $g_{c} = 2\sqrt{2\pi}$ ;
\begin{eqnarray}
ds^2 = {2M\over r}dt^{2} - {r\over 2M}dr^{2}
+ r^{2}d\Omega^{2}_{2}. \nonumber
\end{eqnarray} 
Again this metric represents a spacetime in which $r$ is timelike and
$t$ is spacelike. Also $r = 0$ is a naked singularity with no event
horizon whatsoever around it and hence leads to the violation of the
cosmic censorship hypothesis.
\\
(iii) In the strong coupling limit  ($g_{c} >> 2\sqrt{2\pi}$) ;
\begin{eqnarray}
ds^2 = - [(1 - {8\pi\over g^{2}_{c}}) - {2M\over r}]dt^{2}
+ [(1 - {8\pi\over g^{2}_{c}}) - {2M\over r}]^{-1}dr^{2}
+ r^{2}d\Omega^{2}_{2}. \nonumber
\end{eqnarray}
This metric describes a black hole spacetime with an event horizon
placed at $r = 2M(1 - {8\pi\over g^{2}_{c}})^{-1}$ which encloses
a spacelike curvature singularity at $r = 0$. Since this metric
is characterized by two parameters, $M$ and the non-abelian gauge
coupling constant $g_{c}$, the black hole has a non-abelian hair.
This black hole spacetime is, as emphasized, not merely a coordinate
transformation of the usual Schwarzschild black hole but they have
analogous global structures and thermodynamic properties. For 
instance,
this black hole has Hawking temperature and entropy of $T_{H} = 
(1-{8\pi\over g^{2}_{c}})^2/8\pi M$
and $S = 4\pi M^{2}(1-{8\pi\over g^{2}_{c}})^{-2}$ respectively.
\\
Now we conclude with few observations.
Firstly, the ``non-abelian" hair of the black hole solution in the strong 
coupling
limit possesses an exotic property. Unlike the abelian gauge charge in
the familiar Einstein-Maxwell theory, the non-abelian gauge coupling
parameter $g_{c}$ that characterizes the black hole solution above is {\it 
not} measurable as surface integrals at spatial infinity. This is because
the metric solution is coupled to the vacuum gauge solution $F_{\mu\nu} = 0$
in this EAT theory.
Secondly, the metric solutions for cases (i) and (ii) are shown to 
exhibit the violation of cosmic censorship hypothesis.
They, in fact, provide non-trivial counter-examples to the hypothesis 
in the sense that both the physical and the mathmatical versions of
the hypothesis are violated. Namely, its classical metric solution
turns out to violate the hypothesis while the EAT system itself satisfies
the dominant energy condition (i.e., the locally non-negative matter
energy density), $T_{\mu\nu}n^{\mu}n^{\nu} = 1/g^{2}_{c}r^{2} \geq 0$
(where $n^{\mu}$ is the timelike unit vector)
on which the mathmatical version of the hypothesis is based.
 Note that the cosmic censorship hypothesis is believed to hold in 
the classical theory of general
relativity. And thus far there has been no known concrete example of 
the violation of the
hypothesis with its origin being at the classical theory. ``White holes", whose
existence has been proposed to be possible, should not be regarded as a counter
-example to the ``classical" cosmic censorship conjecture since
they are objects that can be speculated to exist via the
``time-reversal" of the classical black holes in the conventional definition or
 the quantum black holes that do evaporate in Hawking's option [7]. 
In this sense,
our classical metric solution in EAT theory 
appears to be an interesting example that violates the hypothesis in the
purely classical regime. 
It seems, however, fair to point out that the sort of the violation of the
cosmic censorship hypothesis we found here is rather a peculiar consequence
of the ``exotic" classical metric solution that arises when a classical
matter field theory is coupled to Einstein gravity than a phenomenologically
realistic result arising from the gravitational collapse of some well-defined
initial data.

\vspace{2cm}

{\bf \large References}

\begin{description}

\item {[1]} M.Kalb and P. Ramond, Phy. Rev. {\bf D9}, 2273 (1974)
\item {[2]} E. Cremmer and J. Scherk, Nucl. Phys. {\bf B72}, 117 (1974) 
\item {[3]} E. Sezgin and P. van Nieuwenhuizen, Phys. Rev. {\bf D22}, 301 (1980)
\item {[4]} D. Z. Freedman and P. K. Townsend, Nucl. Phys. {\bf B177}, 282 (1981)
\item {[5]} S. P. de Alwis et. al., Nucl. Phys. {\bf B303}, 57 (1988);
A. A. Slavnov and S. A. Frolov, Padova preprint {\bf DFPD} 13/87.
\item {[6]} See, for instance, R. Rajaraman, Solitons and instantons
(North-Holland Publishing company, Amsterdam, 1982) 
\item {[7]} S. W. Hawking, Phys. Rev. {\bf D13}, 191 (1976)

\end{description}

\end{document}